\documentclass[aps,prd,superscriptaddress,10pt]{revtex4-2}
\usepackage{color}
\usepackage{xcolor}
\usepackage{graphicx} 
\makeatletter
\let\PrismOrigPackageWarning\PackageWarning
\def\PrismNamerefPkg{nameref}
\def\PackageWarning#1#2{%
  \def\PrismPkgName{#1}%
  \ifx\PrismPkgName\PrismNamerefPkg\else
    \PrismOrigPackageWarning{#1}{#2}%
  \fi
}
\usepackage{nameref}
\let\PackageWarning\PrismOrigPackageWarning
\makeatother
\usepackage[normalem]{ulem}
\usepackage{amsmath}
\usepackage{amssymb}
\usepackage{float}
\usepackage{comment}
\usepackage{slashed}
\usepackage{silence}

\definecolor{darkred}{rgb}{0.6,0,0}
\usepackage[colorlinks=true,linkcolor=darkred,citecolor=darkred,urlcolor=darkred, pdfborder={0 0 0}]{hyperref}
\usepackage{tikz}
\usepackage[compat=1.1.0]{tikz-feynman}
\tikzfeynmanset{warn luatex=false}

\usepackage{orcidlink}

\begin{document}

\title{Insights into 1-loop corrections to neutrino low-scale type-I seesaw mechanism
}

\author{Gennaro Miele~\orcidlink{0000-0002-2028-0578}}\email{miele@na.infn.it}
\affiliation{Dipartimento di Fisica ``Ettore Pancini'', Università degli studi di Napoli ``Federico II'', Complesso Univ. Monte S. Angelo, I-80126 Napoli, Italy}
\affiliation{INFN - Sezione di Napoli, Complesso Univ. Monte S. Angelo, I-80126 Napoli, Italy}
\author{Stefano Morisi~\orcidlink{0000-0002-6474-8734}}\email{stefano.morisi@unina.it}
\affiliation{Dipartimento di Fisica ``Ettore Pancini'', Università degli studi di Napoli ``Federico II'', Complesso Univ. Monte S. Angelo, I-80126 Napoli, Italy}
\affiliation{INFN - Sezione di Napoli, Complesso Univ. Monte S. Angelo, I-80126 Napoli, Italy}
\author{Eduardo Peinado~\orcidlink{0000-0003-3803-3353}}\email{epeinado@fisica.unam.mx}
\affiliation{Instituto de Física, Universidad Nacional Autónoma de México
Ciudad de México, C.P. 04510, Mexico}
\author{Kainat Qamar~\orcidlink{0009-0001-8466-1222}}\email{kainatqamar@estudiantes.fisica.unam.mx}
\affiliation{Instituto de Física, Universidad Nacional Autónoma de México
Ciudad de México, C.P. 04510, Mexico}

  \begin{abstract}
The standard type-I seesaw can also be regarded as a low-scale seesaw by using the freedom of the Casas-Ibarra parameterization.  
In this framework, radiative corrections to the neutrino mass matrix can dominate over the tree-level contribution, so that the tree-level Casas-Ibarra construction no longer solves the equation that determines the physical neutrino masses and mixings. We show how the Casas-Ibarra parametrization can be consistently extended to this loop-corrected equation, by reabsorbing the 1-loop corrections into the right-handed neutrino mass matrix, so that the resulting light neutrino mass matrix is consistent with experimental values.
On the other hand, we show that physical processes related to right-handed neutrino propagation, such as heavy neutral lepton searches, do not depend on the 1-loop corrections. Moreover, we show that ${\rm Br}(\mu\to e \gamma)$ provides competitive constraints on the parameter space of heavy neutral lepton search experiments for masses above $100$ GeV.
\end{abstract}

\maketitle
\noindent

\section{Introduction}

Neutrino physics is entering an era of percent-level precision in the determination of oscillation parameters. However, the measured oscillation parameters cannot uniquely determine the lepton Yukawa interactions, namely $Y^\ell$ (in the charged sector) and $Y^\nu$ (in the neutral sector). To address this issue, one can follow two approaches.
The first is a “top-down” perspective, in which one considers a specific lepton-Yukawa texture motivated by an underlying symmetry principle. This approach is very popular in the model-building community. The second is a "bottom-up" approach, in which one starts from experimental data to reconstruct Yukawa interactions. The most efficient method in this direction was proposed long ago by Casas and Ibarra \cite{Casas:2001sr}. They provided a simple parameterization of the neutrino Yukawa matrix as a function of neutrino-oscillation observables used as input parameters. This approach is very powerful for model-independent analyses of specific observables, such as branching ratios of lepton-flavor-violating (LFV) processes or lepton asymmetries that depend directly on the heavy-light neutrino coupling, namely the Yukawa matrix $Y^\nu$. In fact, the original paper \cite{Casas:2001sr} studied $Br(\mu \to e \gamma)$ in the framework of the type-I seesaw mechanism.

Moreover, the Casas-Ibarra parametrization reveals an interesting property of the type-I seesaw mechanism. We know that, by naively applying the seesaw relation, one obtains a Yukawa coupling of order one if right-handed neutrinos have masses around the grand unification scale, $\sim 10^{15}~\text{GeV}$, and of order $10^{-8}$ if their masses are about $1~\text{GeV}$. Therefore, any search for light right-handed neutrinos is expected to be suppressed by the smallness of the Yukawa couplings. In contrast, in the case of the so-called "low-energy" seesaw mechanisms \cite{Boucenna:2014zba}, it is possible to have a large Yukawa interactions (even of the order one) right-handed neutrinos masses at the GeV-scale. 
The Casas-Ibarra parametrization shows that, even in the type-I seesaw scenario, large neutrino Yukawa couplings can be obtained for GeV-scale right-handed neutrino masses. Indeed, by using the complex parameters entering the Casas-Ibarra parametrization, one can obtain results similar to those of low-energy seesaw scenarios, namely large Yukawa couplings with light right-handed neutrino masses \cite{Antusch:2017pkq}. For this reason, we refer to this case as the ``low-scale type-I seesaw mechanism". This feature opens many experimental and phenomenological possibilities, in particular for heavy neutral lepton (HNL) searches; see, for instance, \cite{Abdullahi:2022jlv} for a review and discussion of future experimental sensitivities.
However, the neutrino mass mechanism can be affected by radiative corrections, as shown in \cite{Grimus:2002nk}.
A detailed numerical analysis \cite{AristizabalSierra:2011mn} shows that loop corrections can dominate over the tree-level contribution, so that substituting the tree-level Casas-Ibarra solution into the loop-corrected mass matrix no longer reproduces the input oscillation parameters.
As a consequence, the use of the standard type-I seesaw mechanism as a low-energy neutrino mass framework in heavy neutral lepton searches may appear controversial.

In this work, we focus on this aspect and show that this skepticism against the type-I seesaw as a low mass scale is just a bias. To demonstrate this point, we apply the Casas-Ibarra parametrization to the full mass matrix, including the 1-loop correction. As an application, we provide a detailed study of the interplay between the heavy neutral lepton experimental searches and ${\rm Br}(\mu\to e \gamma)$.  

The paper is organized as follows: Section 2 presents the framework and describes the tree-level and 1-loop contributions to neutrino masses. Section 3 discusses the phenomenological implications of heavy-light neutrino mixing through ${\rm Br}(\mu\to e \gamma)$. Section 4 shows how to reabsorb the loop corrections into the Casas-Ibarra parametrization, and Section 5 contains our conclusions.

\section{The Framework}
\subsection*{Neutrino mass at tree-level}
For simplicity, we consider an extension of the Standard Model with three right-handed neutrinos, $N_1$, $N_2$, and $N_3$. In the $(\nu^c_{L_i}, N_i)$ basis, the $6\times 6$ mass matrix is given by
\begin{equation}
\mathcal{M}^{\text{(0)}}
=
\begin{pmatrix}
0 & {M_D^{\text{(0)}}}^{T} \\
M_D^{\text{(0)}} & M_R
\end{pmatrix},
\end{equation}
where we work in the basis in which $M_R$ is diagonal.
By block diagonalizing the matrix $\mathcal{M}^{\text{(0)}}$, one obtains the $3\times 3$ light neutrino mass matrix
\begin{equation}
m_\nu^{\rm (0)}=-{M_D^{\text{(0)}}}^{T} M_R^{-1}  M_D^{\text{(0)}}.
\end{equation}
This matrix can be diagonalized as 
\begin{equation}
U_{PMNS}^T\, m_\nu^{\rm (0)}\, U_{PMNS}= m_\nu^{diag},
\end{equation} where 
\begin{equation}
m_\nu^{diag}=
\begin{pmatrix}
m_1 & 0 & 0\\
0 & \sqrt{m_1^2+\Delta m^2_{sol}} & 0\\
0 &0&  \sqrt{m_1^2+\Delta m^2_{atm}} \\
\end{pmatrix}.
\end{equation}
The leptonic mixing matrix $U_{\text{PMNS}}$ is parameterized as
\begin{equation}
U_{\text{PMNS}} =
\begin{pmatrix}
1 & 0 & 0 \\
0 & c_{23} & s_{23} \\
0 & -s_{23} & c_{23}
\end{pmatrix}
\begin{pmatrix}
c_{13} & 0 & s_{13} e^{-i\delta_{CP}} \\
0 & 1 & 0 \\
- s_{13} e^{i\delta_{CP}} & 0 & c_{13}
\end{pmatrix}
\begin{pmatrix}
c_{12} & s_{12} & 0 \\
- s_{12} & c_{12} & 0 \\
0 & 0 & 1
\end{pmatrix}
P,
\end{equation}

where $c_{ij}=\cos\theta_{ij}$, $s_{ij}=\sin\theta_{ij}$, and the matrix $P$ contains the Majorana phases
\begin{equation}
P=
\begin{pmatrix}
1 & 0 & 0 \\
0 & e^{i\alpha_1/2} & 0 \\
0 & 0 & e^{i\alpha_2/2}
\end{pmatrix}.
\end{equation}
The oscillation parameters $\Delta m^2_{sol}$, $\Delta m^2_{atm}$, $\theta_{ij}$, and $\delta_{CP}$ are experimentally measured \cite{deSalas:2020pgw,Esteban:2024eli,Capozzi:2025wyn}.

\begin{figure}[H]
	\centering
	\includegraphics[width=0.9\textwidth]{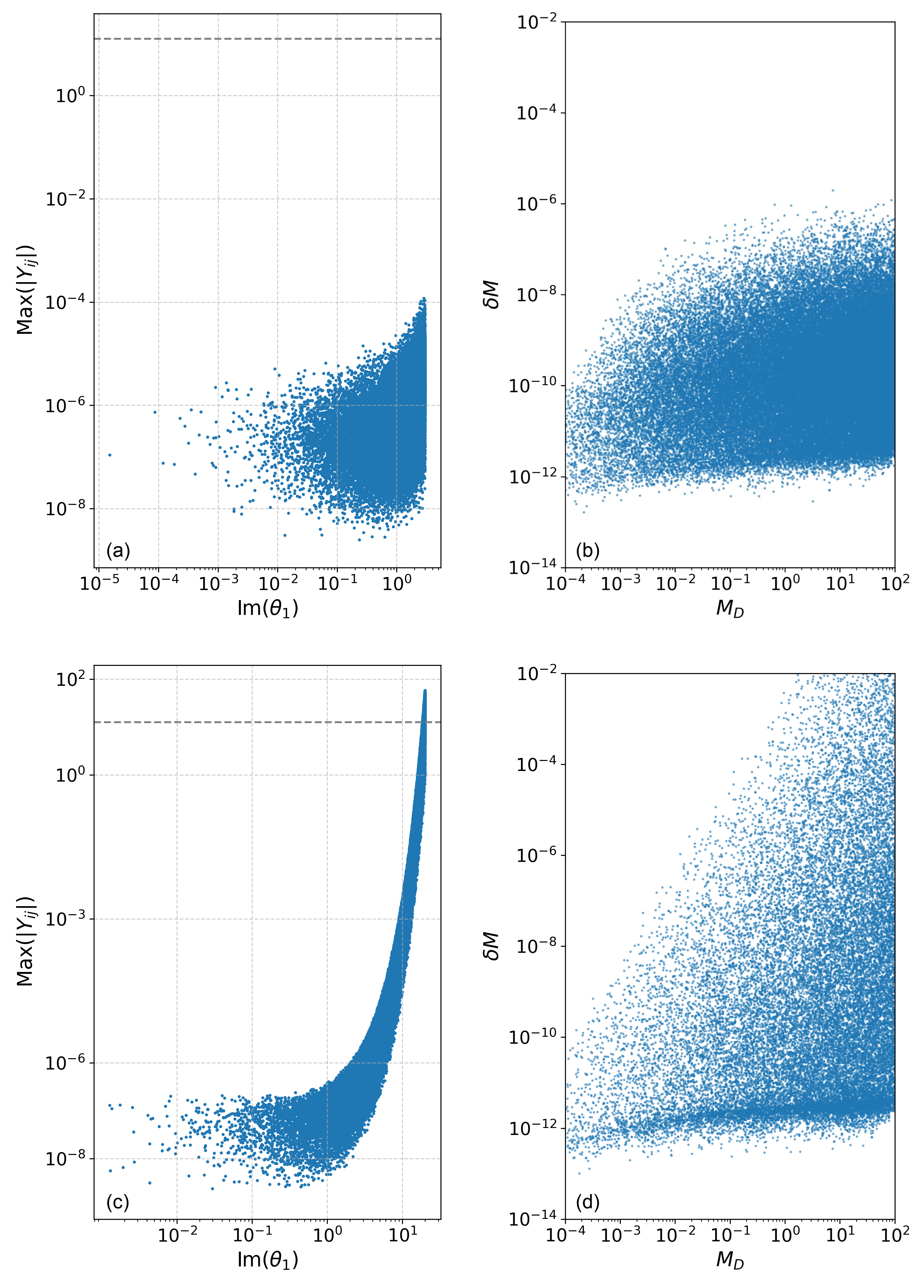}
		\caption{
			Comparison of the maximum Yukawa coupling and the radiative neutrino mass correction for two representative Casas--Ibarra parameter choices.
			Panels (a,b) correspond to $z_1=z_2=z_3=3$, whereas panels (c,d) correspond to $z_1=20$ with $z_2=z_3=0$. 
			The left panels show $\max(|Y_{ij}|)$ as a function of $\mathrm{Im}(\theta_1)$, while the right panels show the radiative mass correction $\delta M$ (GeV) as a function of the Dirac mass scale $M_D$ (GeV).
			The dashed horizontal line denotes the perturbativity limit $\max(|Y_{ij}|)=\sqrt{4\pi}$.
		}
		\label{fig11}

\end{figure}

For numerical analyses, it is useful to express the Dirac neutrino mass matrix $M_D$ in terms of input parameters. This was obtained by Casas and Ibarra in \cite{Casas:2001sr} and is given by
\begin{equation}\label{CI1}
M_D^{\rm (0)}= i \sqrt{M_R} \,R \,\sqrt{m_\nu^{diag}} \,U_{PMNS},
\end{equation} 
where $M_R$ is diagonal and $R$ is a complex orthogonal matrix
\begin{equation}
R =
\begin{pmatrix}
c_2 c_3 &
- c_1 s_3 - s_1 s_2 c_3 &
s_1 s_3 - c_1 s_2 c_3
\\
c_2 s_3 &
c_1 c_3 - s_1 s_2 s_3 &
- s_1 c_3 - c_1 s_2 s_3
\\
s_2 &
s_1 c_2 &
c_1 c_2
\end{pmatrix},
\end{equation}

\begin{equation}
c_i = \cos \theta_i, 
\qquad s_i = \sin \theta_i,
\end{equation}
where the angles $z_i$ ($i=1,2,3$) are complex
\begin{equation}
\theta_i = x_i + i y_i.
\end{equation}
The maximum value of the imaginary part $y_i$ is given by
\begin{equation}
    z_i\equiv |{\rm Max} ({\rm Im} \left(\theta_i)\right)|\,,
\end{equation}
which, as shown below, is the most relevant parameter in the present analysis. 
To satisfy  perturbativity constraint, $\max(|M_D|_{ij})/v<\sqrt{4\pi}$ (equivalently, $|Y_{ij}|^2<4\pi$), the imaginary part of the Casas--Ibarra angle is strongly constrained. As shown in the panel c of Fig.~\ref{fig11}, the perturbative region is predominantly concentrated at $z_i\lesssim 17$, while a small fraction of points satisfying the perturbativity condition can still occur for larger values of $z_i$. 
We define $\delta M$ as the radiative correction to the neutrino mass matrix,
\begin{equation}
m_\nu- m_\nu^{\rm (0)},
\end{equation}
and we quantify its size using the average absolute matrix element,
\begin{equation}
|\delta M| \equiv \frac{1}{9}\sum_{i,j} |\delta M_{ij}|.
\end{equation}
In our numerical scan, only one Casas--Ibarra angle is allowed to have a non-zero imaginary part at a time, while the remaining two are kept real. For Figure (\ref{fig11}), we choose $z_1 \equiv |\mathrm{Im}(\theta_1)| \neq 0$ with $z_2=z_3=0$. This choice reduces the dimensionality of the parameter space while preserving the essential exponential enhancement of the Yukawa couplings induced by complex Casas--Ibarra angles. We explicitly verified that choosing $z_2 \neq 0$ or $z_3 \neq 0$ instead leads to qualitatively similar results with negligible quantitative differences; therefore, only the $z_1$ case is shown for clarity.

%\begin{figure}[H]
%\centering
%\includegraphics[width=0.48\textwidth]{Ymax_vs_Imz1.png}
%\centering
%\includegraphics[width=0.49\textwidth]{deltaM_vs_MD_z1.png}

%\label{fig11-2}
%\end{figure}

\subsection*{Neutrino mass: 1-loop corrections}

The 1-loop corrections to neutrino masses come mainly from the self-energy type diagrams and were computed by Grimus and Lavoura \cite{Grimus:2002nk}. In particular, the authors found that the most relevant correction affects the upper-left sub-block of the neutrino mass matrix. The resulting full neutrino mass matrix is 
\begin{equation}\label{full1loop}
\mathcal{M}^{\text{(1)}}
=
\begin{pmatrix}
\delta M & {M_D^{\text{(0)}}}^{T} \\
{M_D^{\text{(0)}}} & M_R
\end{pmatrix},
\end{equation}
where $\delta M$ denotes the 1-loop contribution and is given by
\begin{equation}
\delta M={M_D^{(0)}}^{T}\,\,M_R^{-1} \,{\Delta_1}\,{M_D^{(0)}},
\end{equation}
where $\Delta_1$ is a diagonal matrix (in the simplest one-Higgs case), given by
\begin{equation}
(\Delta_1)_{ij} =
\frac{g^2}{64 \pi^2 M_W^2\,} \,\left(
m_h^2 \log\left( \frac{M_i^2}{m_h^2} \right)+3 M_Z^2 \log\left( \frac{M_i^2}{M_Z^2} \right)\right)  \,\delta_{ii} ,  
\end{equation}

where $M_i$ are the right-handed masses.
After block-diagonalizing $\mathcal{M}^{\text{(1)}}$, the resulting light-neutrino mass matrix is
\begin{equation}\label{neumass1loop}
 m_\nu=m_\nu^{\rm (0)}+\delta M \, \equiv \,  - {M_D^{\text{(0)}}}^{T} \,M_R^{-1} \, {M_D^{\text{(0)}}} \,+ \, {M_D^{\text{(0)}}}^{T} \,M_R^{-1} \,\,{\Delta_1}\, {M_D^{\text{(0)}}} \,.
\end{equation}

\section{Heavy-light neutrino mixing}

The Casas-Ibarra parametrization is a powerful tool for studying physical processes involving heavy-light neutrino mixing. In this context, it is important to diagonalize the full $6\times 6$ neutrino mass matrix and determine the corresponding mixing matrix, which encodes both neutrino oscillation parameters and heavy-light neutrino mixing.

As an illustration, we consider experiments searching for heavy neutral leptons, such as ANUBIS \cite{Bauer:2019vqk,Shah:2024fpl}, MATHUSLA \cite{Curtin:2018mvb}, SHADOWS \cite{Baldini:2021hfw}, NA62 \cite{NA62:2017rwk,Drewes:2018gkc}, FASER \cite{FASER:2018eoc},
CODEX-b \cite{Gligorov:2017nwh}, SHiP \cite{SHiP:2015vad},
FCC-ee\,\cite{FCC:2018evy}, and FCC-hh\,\cite{FCC:2018vvp}.
These experiments play a crucial role in testing the low-scale type-I seesaw mechanism through the production and decay of heavy neutral leptons (HNLs). The production cross section and decay widths of the heavy neutrinos are governed by their mixing with the active neutrinos. For each lepton flavour, the mixing strength is defined as
\begin{align}
	U_e^2 &\equiv \sum_{i=4}^{6}|U_{ei}|^2,\\
	U_\mu^2 &\equiv \sum_{i=4}^{6}|U_{\mu i}|^2,\\
	U_\tau^2 &\equiv \sum_{i=4}^{6}|U_{\tau i}|^2,
\end{align}
where $U_{\alpha i}$ denotes the mixing between the active neutrino of flavour $\alpha=(e,\mu,\tau)$ and the heavy neutrino mass eigenstate $N_i$ $(i=4,5,6)$. The total active--sterile mixing is then given by
\begin{equation}
	U^2 \equiv U_e^2+U_\mu^2+U_\tau^2
	=\sum_{\alpha=e,\mu,\tau}\sum_{i=4}^{6}|U_{\alpha i}|^2.
\end{equation}
where $U^\dagger {\mathcal{M}^{\text{(1)}}}^\dagger \mathcal{M}^{\text{(1)}}\,U \sim{\rm Diag}$.

To obtain the exact mixing matrix $U$ and the corresponding light and heavy neutrino mass eigenvalues, we consider the following Hermitian matrix:
\begin{equation}
{\mathcal{M}^{\text{(1)}}}^\dagger \mathcal{M}^{\text{(1)}}
\approx
\begin{pmatrix}
\delta M^\dagger \delta M+ M_D^* M_D^\dagger & \delta M^\dagger M_D+M_D^* M_R \\
M_D^\dagger \delta M + M_R M_D^T & M_R M_R
\end{pmatrix}.
\end{equation} 
The mixing between light and heavy neutrinos ($U_{\nu-N}$) is naively of the order
\begin{equation}
    U_{\nu-N}\sim (\delta M^\dagger M_D+M_D^* M_R)\cdot \frac{1}{(M_R^2)}\,.
\end{equation}
We observe that, since $\delta M$ is of the order of light neutrino masses, one expects $\delta M\ll M_D \ll M_R$ (see the right panels of Figure~(\ref{fig11}).

Therefore, the mixing between light and heavy neutrinos, $U_{\nu-N}$, is not significantly affected by loop corrections:
\begin{equation}
    U_{\nu-N}\sim (M_D^* M_R)\cdot \frac{1}{(M_R^2)}\,.
\end{equation}

Therefore, all observables that mainly depend on light-heavy neutrino mixing (such as $Br(\mu\to e \gamma)$ and $U^2$) are not modified by loop corrections. 
In contrast, the light neutrino mass is strongly affected by loop corrections, which is why the Casas-Ibarra parametrization must be applied to the full loop-corrected mass matrix.

Following \cite{Morisi:2024yxi}, we present the projected experimental exclusion limits in the $(M_i,U^2)$ plane together with the constraints from ${\rm Br}(\mu\to e \gamma)$. The interplay between the process $\mu \to e \gamma$ and heavy neutral lepton searches has also been extensively studied in \cite{Drewes:2016jae, Drewes:2015iva, Chrzaszcz:2019inj, Agrawal:2021dbo, Granelli:2022eru}.

\begin{figure}[H]  
\centering
\includegraphics[width=0.48\textwidth]{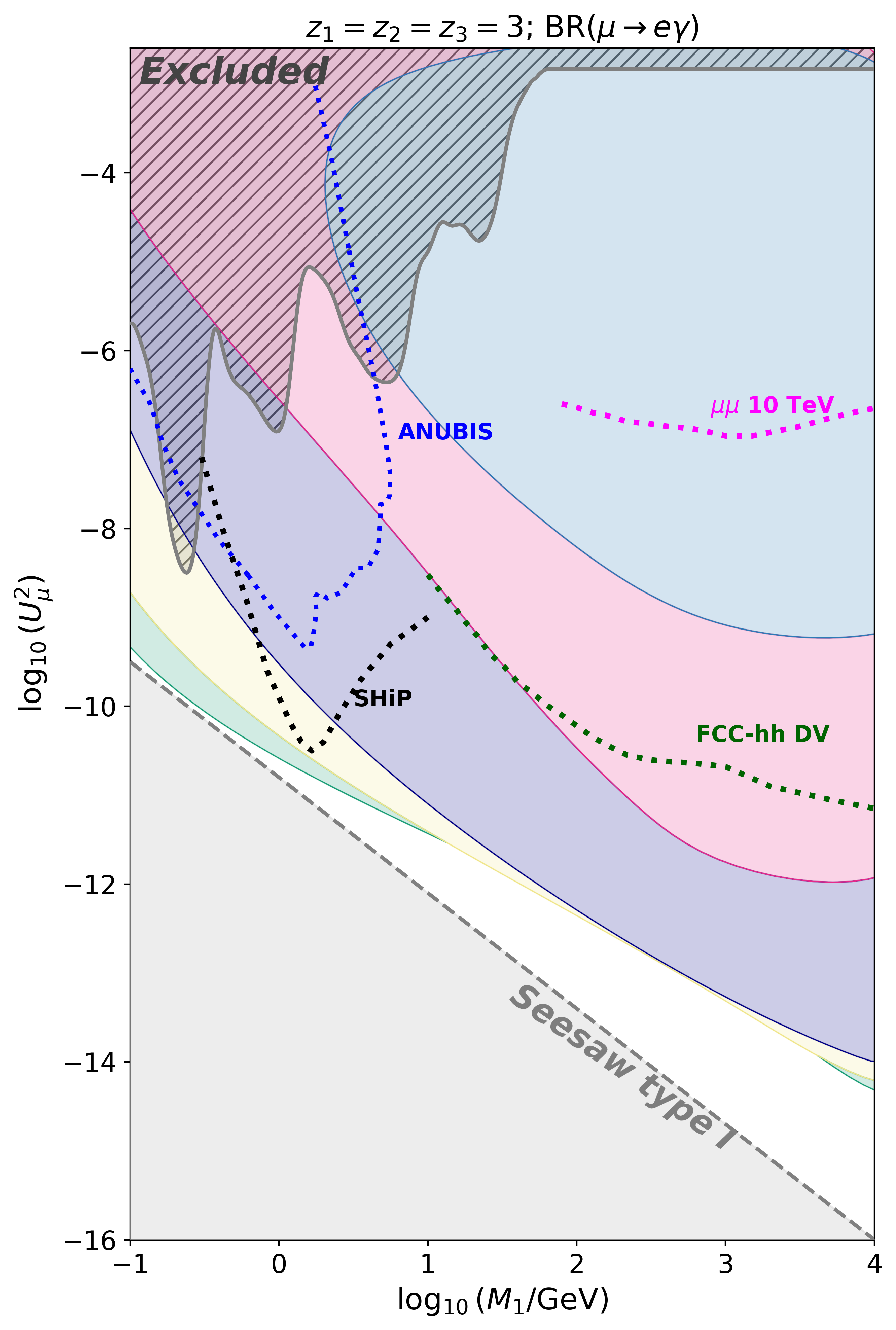}
\includegraphics[width=0.48\textwidth]{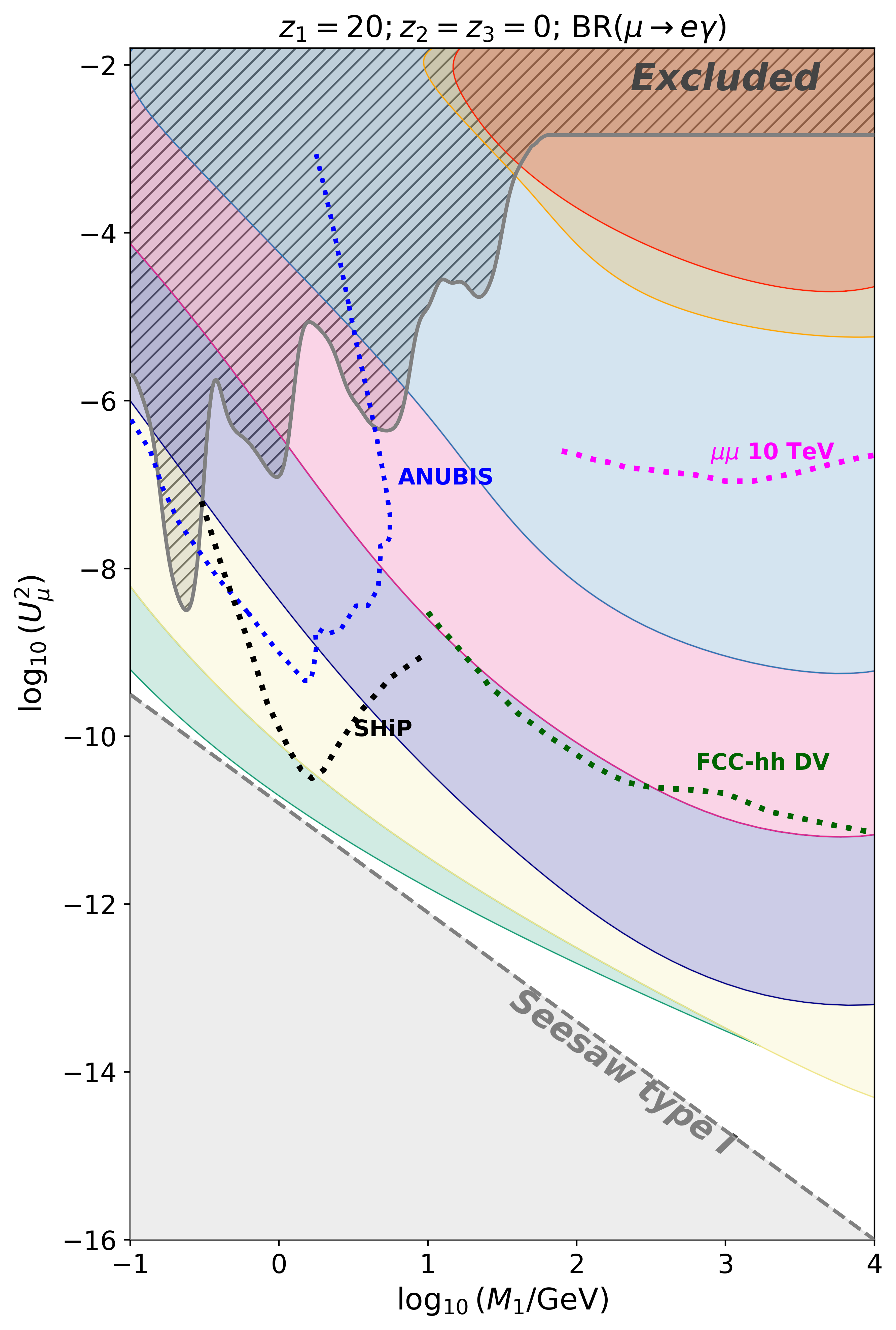}
\caption{$U^2$ as a function of $M_1$. The parameter space is color-coded according to the branching ratio: green ($\mathrm{Br} > 10^{-35}$), khaki ($\mathrm{Br} > 10^{-32}$), purple ($\mathrm{Br} > 10^{-30}$), pink ($\mathrm{Br} > 10^{-26}$), blue ($\mathrm{Br} > 10^{-22}$), orange ($\mathrm{Br} > 10^{-14}$), and red ($\mathrm{Br} > 10^{-13}$). The additional curves indicate projected experimental sensitivities (see text for details).}
\label{figship1}
\end{figure}

Our results are shown in Figure~(\ref{figship1}) for different choices of the parameters $z_i$, where
$U_\mu^2$ measures the total mixing of the muon flavor with the heavy neutrino states. The color coded zones (The points are obtained from a full random scan over the model parameter space, including the light-neutrino oscillation parameters, the Casas--Ibarra angles, and the heavy Majorana masses. No heavy-neutrino mass or mixing parameter is held fixed.) represent different lower bounds on $Br(\mu\to e \gamma)$. In particular, the latest MEG II result is shown in red, corresponding to $Br(\mu\to e \gamma)>1.5 \times 10^{-13}(90\% \,{\rm C.L.})$ \cite{MEGII:2025gzr}.
This bound currently provides the most stringent constraint, even when compared with those from the LHC.

\begin{figure}[H]  
\centering
\includegraphics[width=0.6\textwidth]{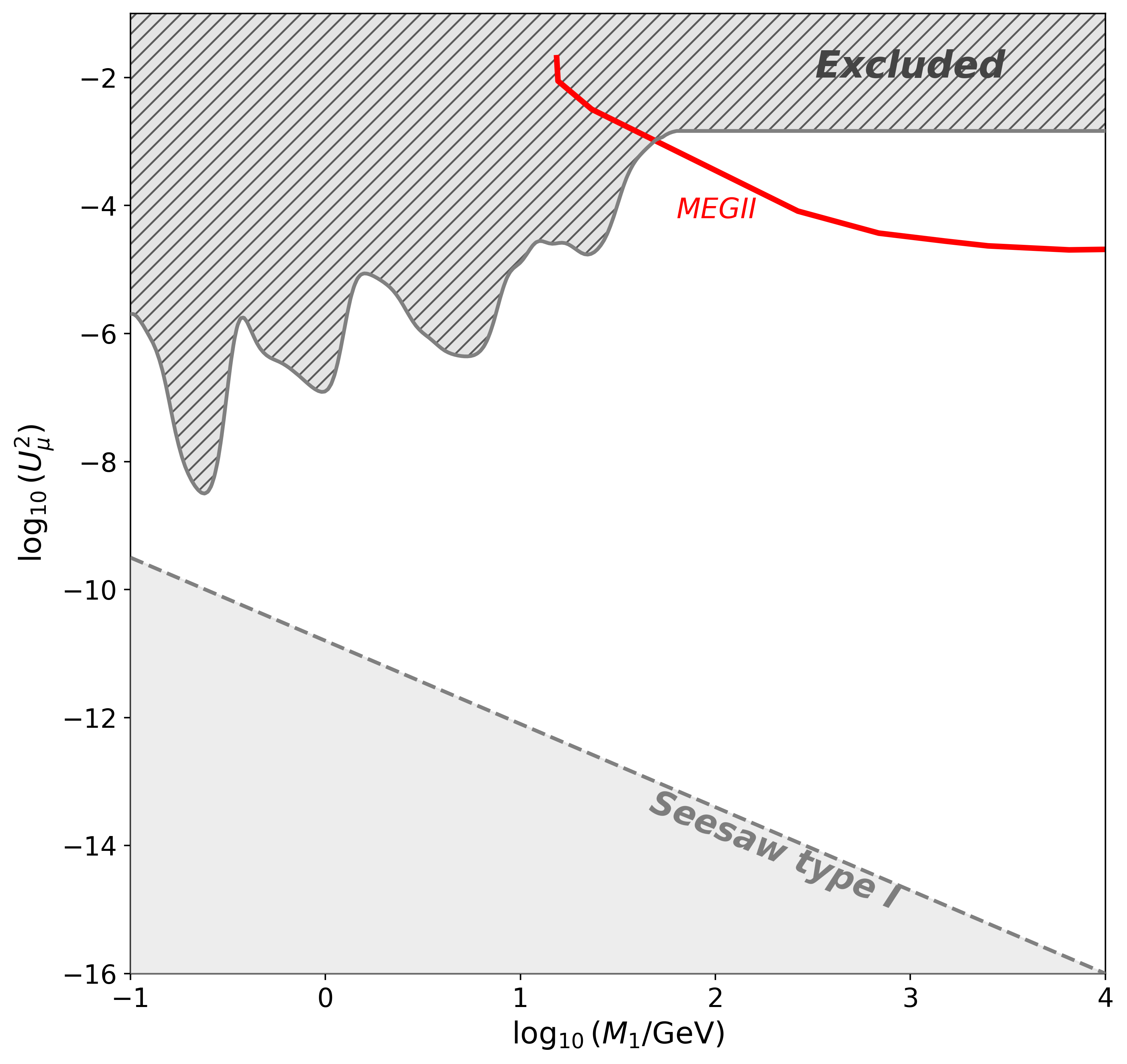}
\caption{Summary plot of $\mu\to e \gamma$ found in this paper.}\label{figsummary}
\end{figure}

We use the updated experimental constraints on heavy neutral lepton searches reported in~\cite{Abdullahi:2022jlv} to summarize our results on $\mu\to e \gamma$ limits in Figure~(\ref{figsummary}).

Moreover, we note that, in the case of a positive detection in heavy neutral lepton experiments, this discovery can be attributed to a right-handed neutrino involved in the type-I seesaw mechanism. In this case, we can infer a lower bound on ${\rm Br}(\mu\to e \gamma)$ in the range of about $10^{-35}\text{--}10^{-14}$. This bound is significantly enhanced with respect to the Standard Model value of $10^{-54}$.

\begin{figure}[H]  
\centering
\includegraphics[width=0.45\textwidth]{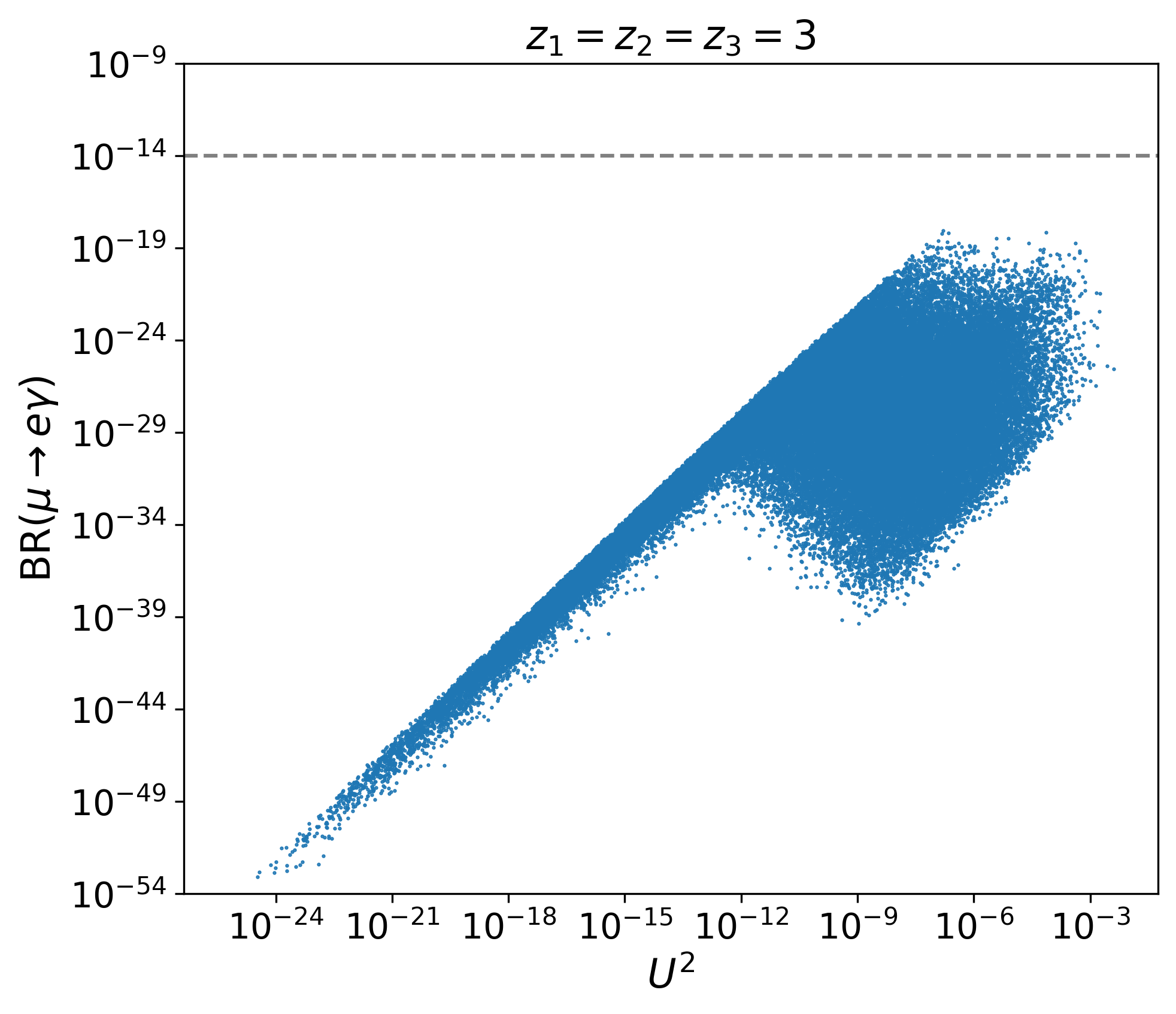}
\includegraphics[width=0.45\textwidth]{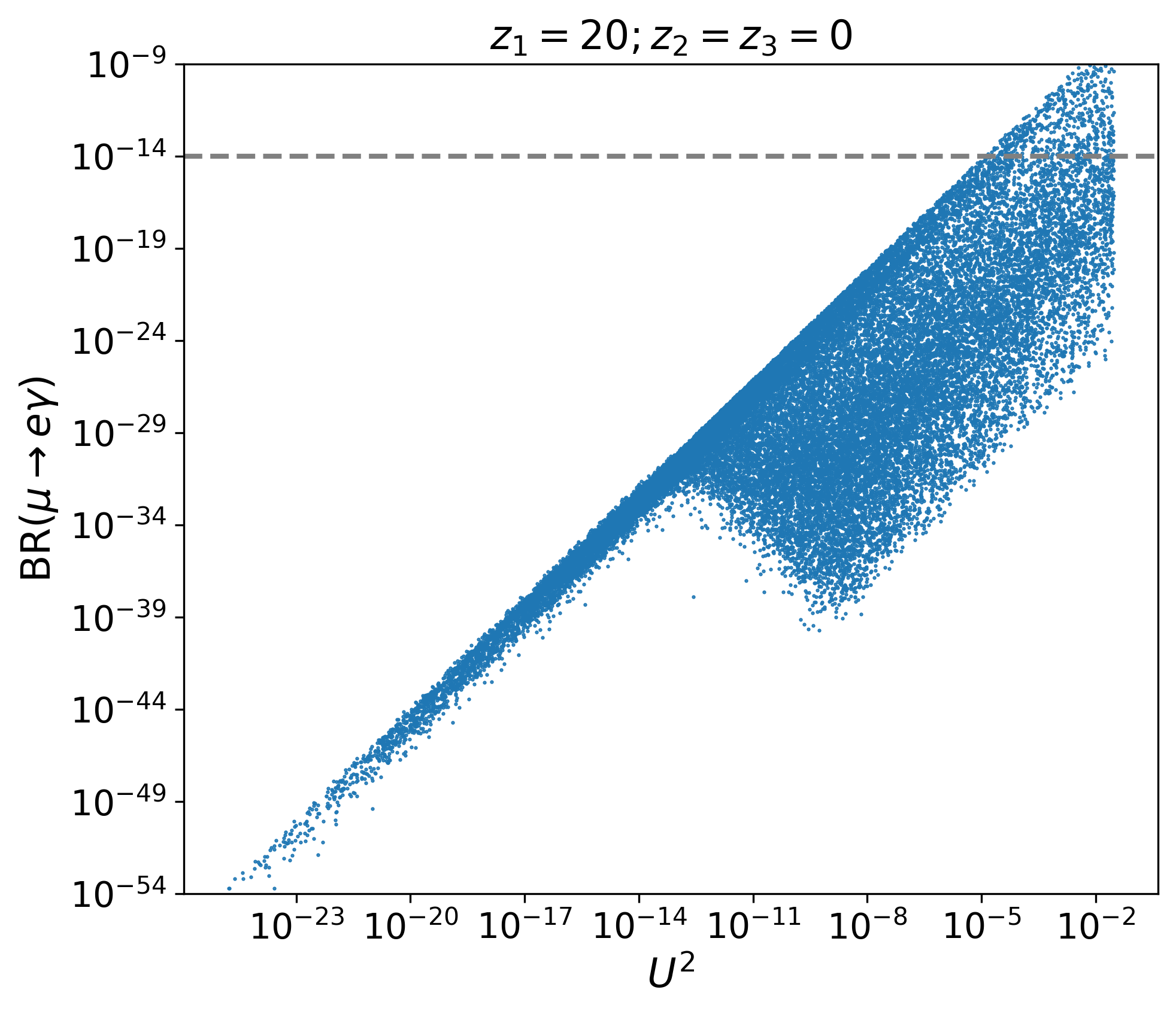}

\caption{$U^2$ versus $BR(\mu\to e \gamma)$ for two different choices of complex angles, 
The horizontal dashed line indicates the current experimental upper bound. Larger imaginary Casas--Ibarra angles enhance the lepton-flavor-violating branching ratio.}\label{figship2}
\end{figure}

It is also useful to examine the relationship between the branching ratio of $\mu\to e \gamma$ and the mixing parameter $U^2$, as shown in Figure~(\ref{figship2}).

\section{Including loop corrections within the Casas-Ibarra parametrization}

\subsection*{The 1-loop case}

In the presence of 1-loop corrections, the light neutrino mass matrix is given by the full expression in Eq.~(\ref{neumass1loop}), rather than by the tree-level relation alone. We note that the Casas-Ibarra parametrization can be consistently extended to this loop-corrected case, as discussed in \cite{Lopez-Pavon:2015cga}.
In this framework, the light neutrino mass matrix can be written as 
\begin{equation}
m_\nu=m_\nu^{\rm (0)}+\delta M \equiv - {M_D^{\text{(1)}}}^{T} \frac{1}{M_R^\prime} {M_D^{\text{(1)}}}\,.
\end{equation}
where 
\begin{equation}\label{MN2}
\frac{1}{M_R^\prime} = M_R^{-1}\left(1-\Delta_1\right).
\end{equation}

\vspace{0.5cm}

Therefore, the Casas-Ibarra parametrization follows in a similar way to the tree-level case with the replacement of $M_R$ with $M_R^\prime$ defined in (\ref{MN2}) and we simply have
\begin{equation}\label{CI2}
M_D^{\text{(1)}} = i \sqrt{M_R^{\prime}}\, R \,\sqrt{m_\nu^{diag}} \, U_{PMNS}.
\end{equation} 

\noindent
Then, instead of (\ref{full1loop})  we assume   for 
$\mathcal{M}^{\text{(1)}}$   the following expression 
\begin{equation}\label{full1loopp}
\mathcal{M'}^{\text{(1)}}
=
\begin{pmatrix}
\delta M'& {M_D^{\text{(1)}}}^{T} \\
{M_D^{\text{(1)}}} & M_R
\end{pmatrix},
\end{equation}
where $ \delta M'= {M_D^{\text{(1)}}}^{T}\, M_R^{-1}\,{\Delta_1} \,{M_D^{\text{(1)}}}$ and ${M_D^{\text{(1)}}}$ are defined in (\ref{CI2}).

\begin{figure}[H]  
\centering
\includegraphics[width=0.47\textwidth]{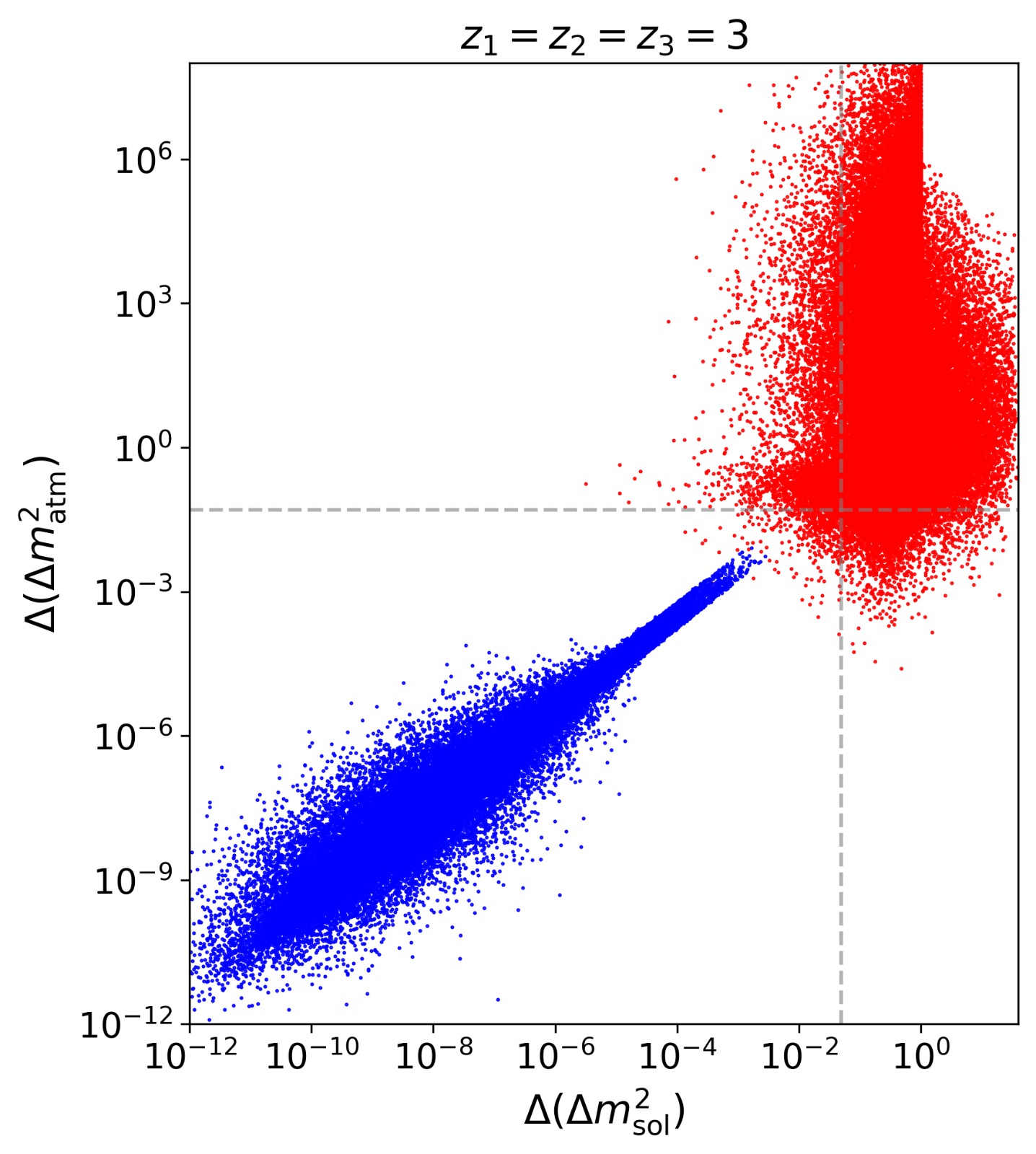}
\includegraphics[width=0.47\textwidth]{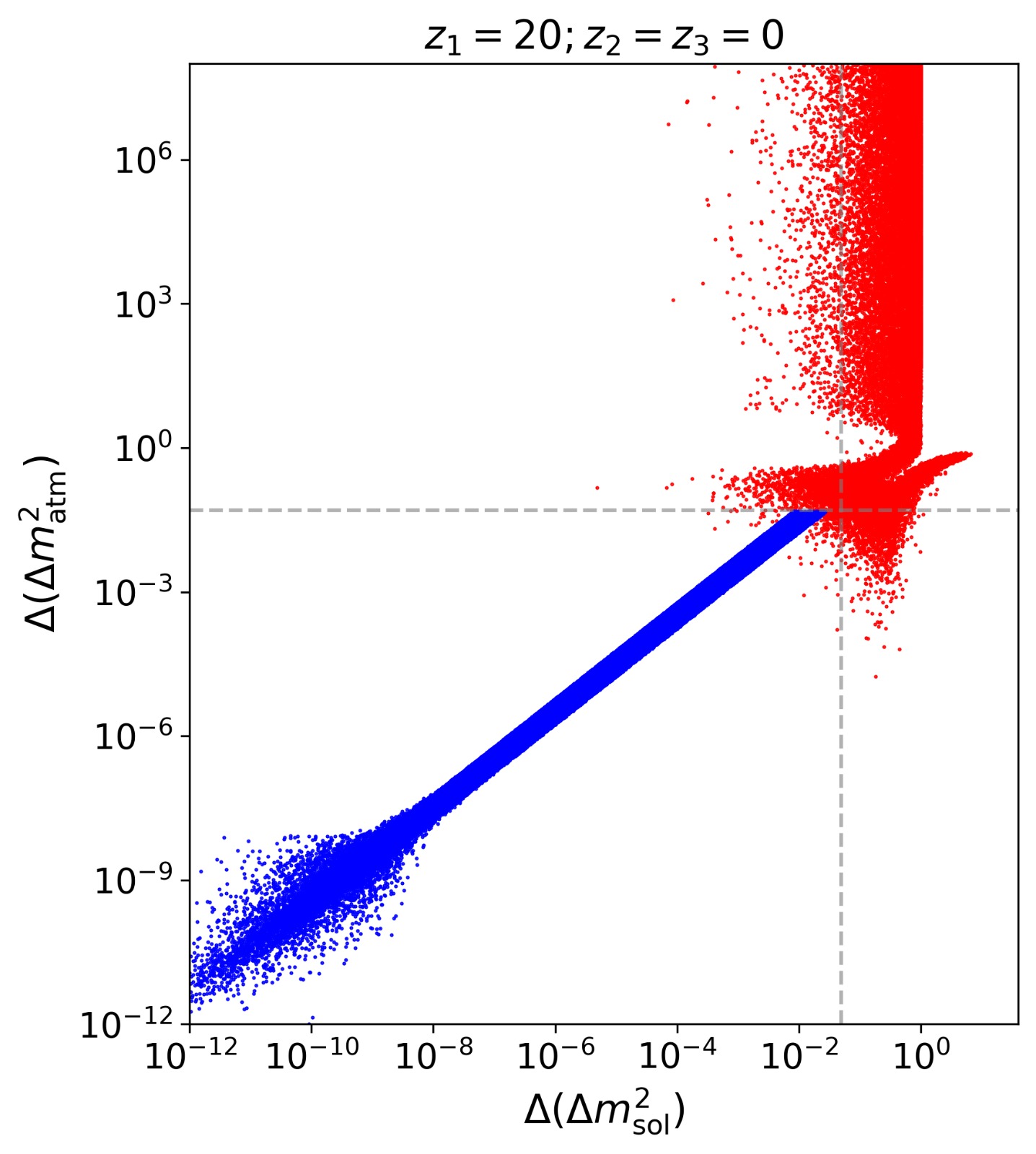}
\caption{Variation of the neutrino mass-squared differences obtained using Eq.~(\ref{CI1}) (red points) and Eq.~(\ref{CI2}) (blue points), including dominant loop corrections relative to the best-fit values for different choices of complex angles, see the text for details. }\label{figDeltam}
\end{figure}

\begin{figure}[H]  
\centering
\includegraphics[width=0.47\textwidth]
{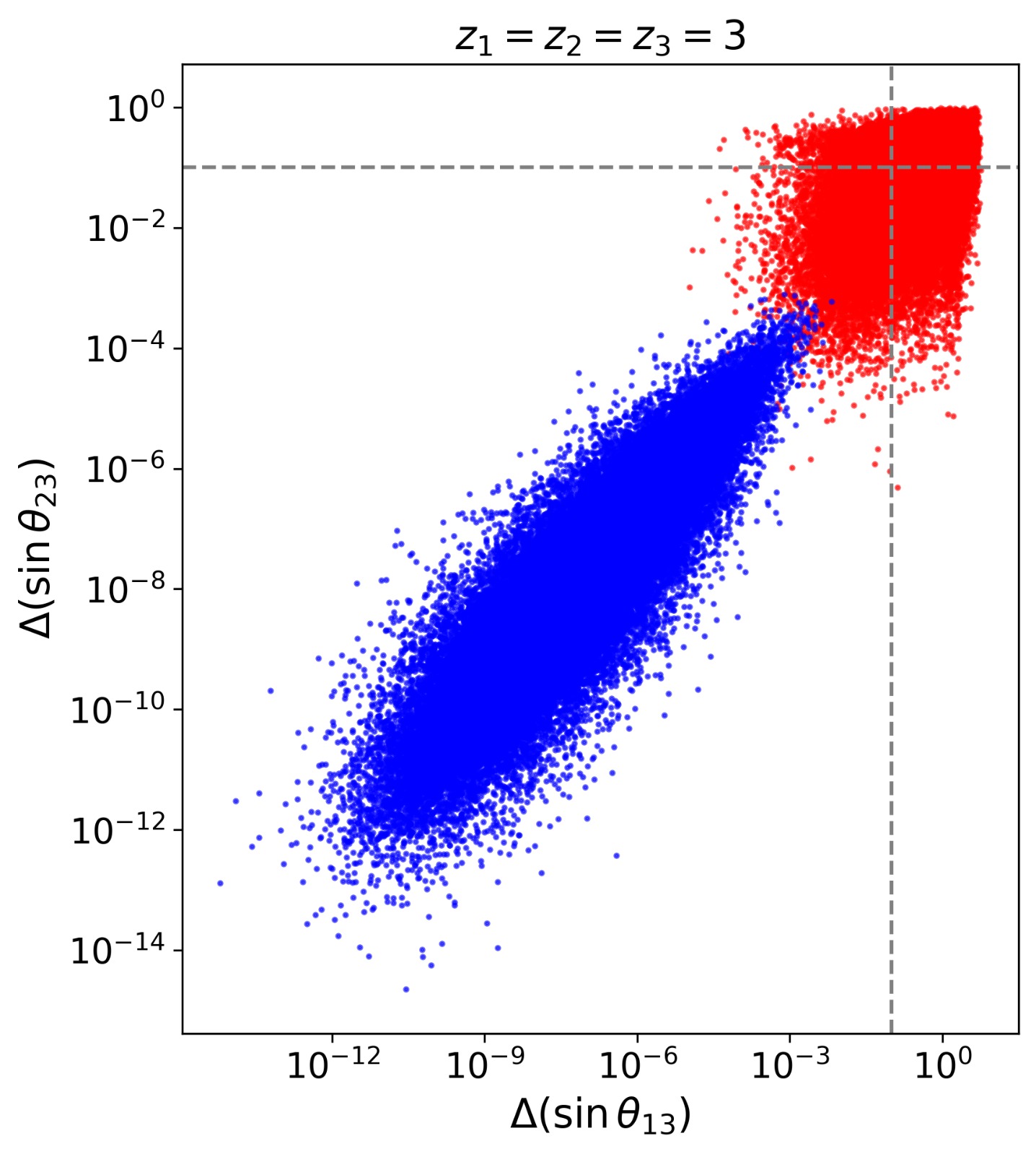}
\includegraphics[width=0.47\textwidth]
{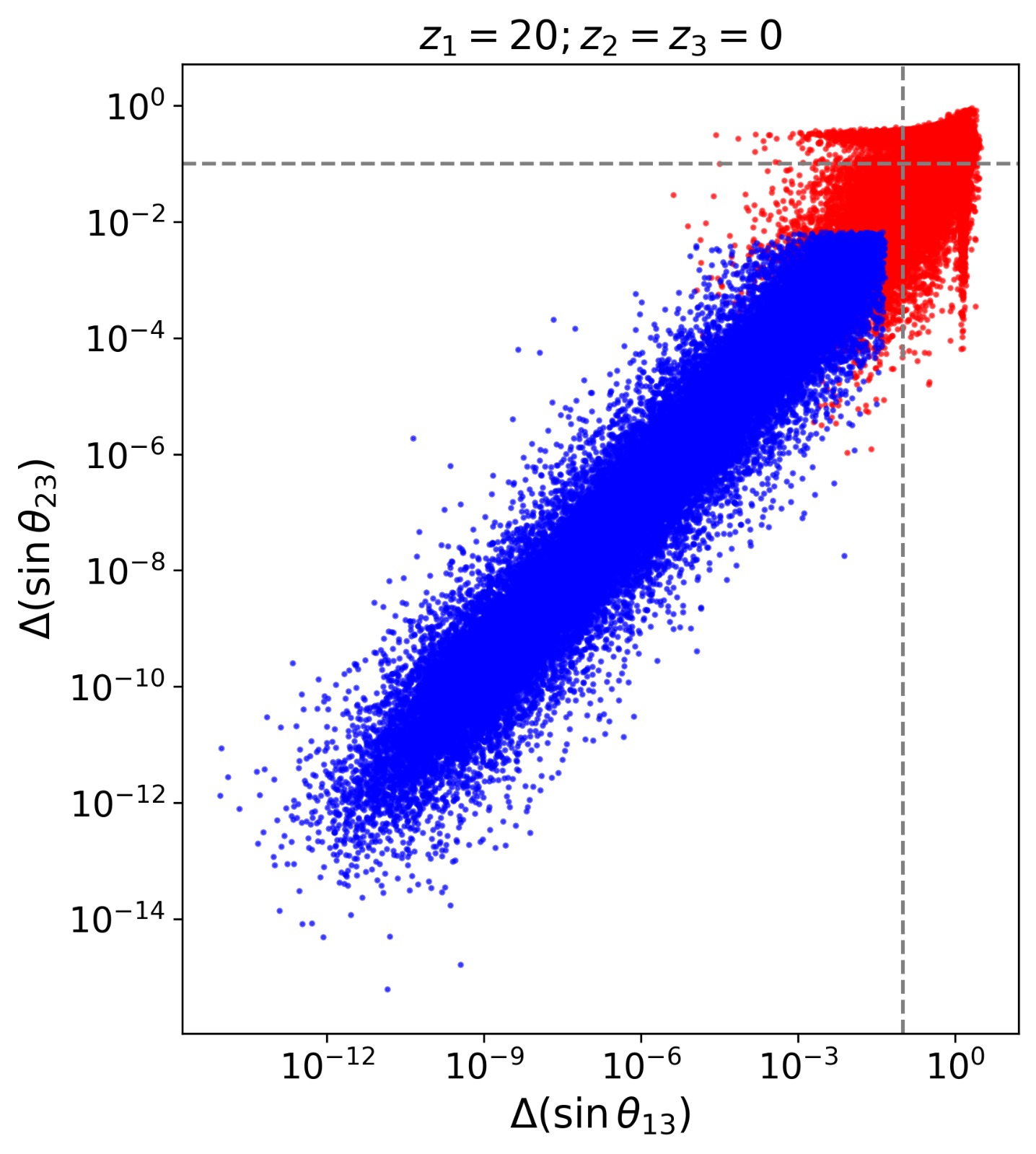}
\caption{Variation of the neutrino mixing angle differences obtained using Eq.~(\ref{CI1}) (red points) and Eq.~(\ref{CI2}) (blue points), including dominant loop corrections relative to the best-fit values for different choices of complex angles.}\label{figDeltatheta}
\end{figure}

Using $\mathcal{M'}^{\text{(1)}}$ instead of $\mathcal{M}^{\text{(1)}}$, and diagonalizing numerically ${\mathcal{M'}^{\text{(1)}}}^\dagger \mathcal{M'}^{\text{(1)}}$, the resulting neutrino oscillation parameters are in perfect agreement with the experimental values, as shown in Figures\,(\ref{figDeltam}) and\,(\ref{figDeltatheta}) with blue points. This shows that the Casas-Ibarra parametrization (\ref{CI2}) provides correct results for the neutrino oscillation parameters, whereas Eq.~(\ref{CI1}), which solves only the tree-level relation, does not reproduce them once the 1-loop term is included (red points).

Thus, in the presence of 1-loop corrections, the Casas-Ibarra parametrization must be applied to the loop-corrected mass matrix, which gives Eq.~(\ref{CI2}).
We comment on the possible impact of higher–order
radiative corrections beyond the dominant 1-loop contribution
discussed above.  A complete computation of the neutrino
mass matrix at two or more loops is currently not available.
However, it is possible to estimate the size of such effects
by exploiting the loop–suppression structure of the radiative
corrections.

Higher–order loop contributions are expected to be suppressed
by additional powers of the loop factor $(16\pi^2)^{-1}$.
Following this observation, we estimate the structure of the
next corrections by introducing additional loop suppression
factors multiplying the same operator structure appearing in
$\delta M'$. For more details, see the appendix. The result is shown in Figure~(\ref{fig111}) where we estimate the effect of 2- and 3-loop corrections, and we show that the method seems to be convergent.  Namely, going to higher loop corrections the difference between the extended and tree-level constructions becomes smaller and smaller. Of course, this is just an estimation and a full 2- and 3-loop calculation is required but this is far from the scope of this work.

\section{Conclusion}
It is well known that the natural scale of the type-I seesaw mechanism lies close to the grand unification scale, $\sim 10^{15}$\,GeV. If the right-handed neutrinos have masses at this scale and the neutrino Yukawa couplings are of order one, the resulting light-neutrino masses are naturally of order $0.1$\,eV, in agreement with experimental observations.

On the other hand, it is not forbidden to consider right-handed neutrinos at much lower scales, such as the GeV scale (or even below). In this case, one would naively expect the corresponding neutrino Yukawa couplings to be strongly suppressed, rendering any associated collider phenomenology negligible.
For this reason, low-energy seesaw mechanisms, such as inverse or linear seesaw (as well as radiative scenarios such as scotogenic models), have become very popular. However, this is not the only possibility. Indeed, by exploiting the complex parameters of the Casas-Ibarra parametrization, it is possible to enhance the neutrino Yukawa couplings (and the related phenomenology) within the framework of the standard type-I seesaw.

Motivated by this observation, several experimental proposals aim to search for heavy neutral leptons (such as right-handed neutrinos) and probe regions of the standard type-I seesaw parameter space. The general strategy of such experiments (e.g., SHiP and ANUBIS) is to produce a right-handed neutrino on-shell, which can propagate through the detector before decaying. The rates of these processes are governed by the neutrino Yukawa interactions and are proportional to the active-sterile neutrino mixing parameter $U^2$. Experimental sensitivities are therefore typically presented in the $(U^2,M_1)$ plane, where $M_1$ denotes the mass of the lightest right-handed neutrino.
In this work, we show that the current MEG II limits on ${\rm Br}(\mu\to e \gamma)$ can be used to derive new, competitive constraints in the $(U^2,M_1)$ plane for $M_1>100$\,GeV.

Moreover, we point out that, in the event of a positive signal in heavy neutral lepton searches, such a discovery could be interpreted as evidence for a right-handed neutrino participating in the type-I seesaw mechanism. In this scenario, one can infer a lower bound on ${\rm Br}(\mu\to e \gamma)$ in the range $10^{-35}\text{--}10^{-14}$.

We then investigate the impact of 1-loop corrections in the context of such a low-energy type-I seesaw mechanism. We find that the effect of 1-loop corrections on collider phenomenology (such as processes related to heavy neutral lepton searches) is suppressed, even when loop corrections are relevant for light-neutrino masses. Furthermore, we show that the Casas-Ibarra parametrization can be consistently extended to the loop-corrected neutrino mass matrix by reabsorbing the 1-loop corrections into the right-handed neutrino mass matrix, so that the resulting construction reproduces the correct neutrino oscillation parameters.

\section*{acknowledgments}
This work was supported by DGAPA UNAM Grant No. PAPIIT-IN111625, SECIHTI Project No.
CBF-2025-I-1589 and Fundaci\'on Marcos Moshinsky. We thank Avelino Vicente for useful discussions.

    \appendix

\section{Effect of higher order corrections}

Here, we present a detailed discussion of how information about higher-order radiative corrections to neutrino masses can be obtained. Our analysis is intended as a preliminary study, since a complete 2-loop computation is currently unavailable and lies beyond the scope of the present work. Nevertheless, it is useful to estimate the expected size of next-to-leading-order corrections in order to evaluate the convergence of the perturbative expansion.

We note that higher-order loop contributions are expected to be suppressed
by additional powers of the loop factor $(16\pi^2)^{-1}$.
Following this observation, we estimate the structure of the
next corrections by introducing additional loop-suppression
factors multiplying the same operator structure that appears in
$\delta M'$.
Schematically, the one-, two-, and three-loop contributions can be written as\\
\noindent {\bf 1-loop}
\begin{equation}
\mathcal{M}^{\text{(1)}}
=
\begin{pmatrix}
\delta M_1& {M_D^{\text{(0)}}}^{T} \\
{M_D^{\text{(0)}}} & M_R
\end{pmatrix}\,,\quad
\mathcal{M'}^{\text{(1)}}
=
\begin{pmatrix}
\delta M'_1& {M_D^{\text{(1)}}}^{T} \\
{M_D^{\text{(1)}}} & M_R
\end{pmatrix},
\end{equation}
with
\begin{equation}
\delta M_1={M_D^{\text{(0)}}}^{T} \, M_R^{-1}{\Delta_1} \, {M_D^{\text{(0)}}}\,,\quad
\delta M'_1={M_D^{\text{(1)}}}^{T} \, M_R^{-1} {\Delta_1}  \, {M_D^{\text{(1)}}}\,.
\end{equation}
\noindent {\bf 2-loop}
\begin{equation}
\mathcal{M}^{\text{(2)}}
=
\begin{pmatrix}
\delta M_1'+\delta M_2& {M_D^{\text{(1)}}}^{T} \\
{M_D^{\text{(1)}}} & M_R
\end{pmatrix}\,,\quad
\mathcal{M'}^{\text{(2)}}
=
\begin{pmatrix}
\delta M'_1+\delta M'_2& {M_D^{\text{(2)}}}^{T} \\
{M_D^{\text{(2)}}} & M_R
\end{pmatrix},
\end{equation}
with
\begin{equation}
\delta M_2={M_D^{\text{(1)}}}^{T} \, M_R^{-1} {\Delta_2}\, {M_D^{\text{(1)}}}\,,\quad
\delta M'_2={M_D^{\text{(2)}}}^{T} \, M_R^{-1}{\Delta_2} \, {M_D^{\text{(2)}}}\,,
\end{equation}
\begin{equation}
M_D^{\text{(2)}} = i \sqrt{M_R^{\prime\prime}}\, R \,\sqrt{m_\nu^{diag}} \, U_{PMNS}\,,\quad
M_R''=(M_R^{-1} (1-{\Delta_1}-{\Delta_2}))^{-1}.
\end{equation} 
\noindent {\bf 3-loop} 
\begin{equation}
\mathcal{M}^{\text{(3)}}
=
\begin{pmatrix}
\delta M_1'+\delta M'_2 +\delta M_3& {M_D^{\text{(2)}}}^{T} \\
{M_D^{\text{(2)}}} & M_R
\end{pmatrix}\,,\quad
\mathcal{M'}^{\text{(3)}}
=
\begin{pmatrix}
\delta M'_1+\delta M'_2+\delta M'_3& {M_D^{\text{(3)}}}^{T} \\
{M_D^{\text{(3+)}}} & M_R
\end{pmatrix},
\end{equation}
with
\begin{equation}
\delta M_3={M_D^{\text{(2)}}}^{T} \, M_R^{-1} {\Delta_3}\, {M_D^{\text{(2)}}}\,,\quad
\delta M'_3={M_D^{\text{(3)}}}^{T} \, M_R^{-1}{\Delta_3} \, {M_D^{\text{(3)}}}\,,
\end{equation}
\begin{equation}
M_D^{\text{(3)}} = i \sqrt{M_R^{\prime\prime\prime}}\, R \,\sqrt{m_\nu^{diag}} \, U_{PMNS}\,,\quad
M_R''=(M_R^{-1} (1-{\Delta_1}-{\Delta_2}-{\Delta_3}))^{-1},
\end{equation} 
where 
\begin{equation}
  {\Delta_2}=\frac{1}{16 \pi^2}{\Delta_1}\,,\qquad{\Delta_3}=\frac{1}{(16 \pi^2)^2}{\Delta_1}\,.
\end{equation}

These contributions are successively incorporated into the
full neutrino mass matrix used in the numerical analysis.
For each order ($n$) of the expansion, the complete $6\times6$
matrices $\mathcal{M}^{\text{(n)}}$ and $\mathcal{M'}^{\text{(n)}}$ are diagonalized numerically, and
the resulting light neutrino spectrum is used to reconstruct
the solar and atmospheric mass–squared differences.

\begin{figure}[H]  
\centering
\includegraphics[width=0.5\textwidth]{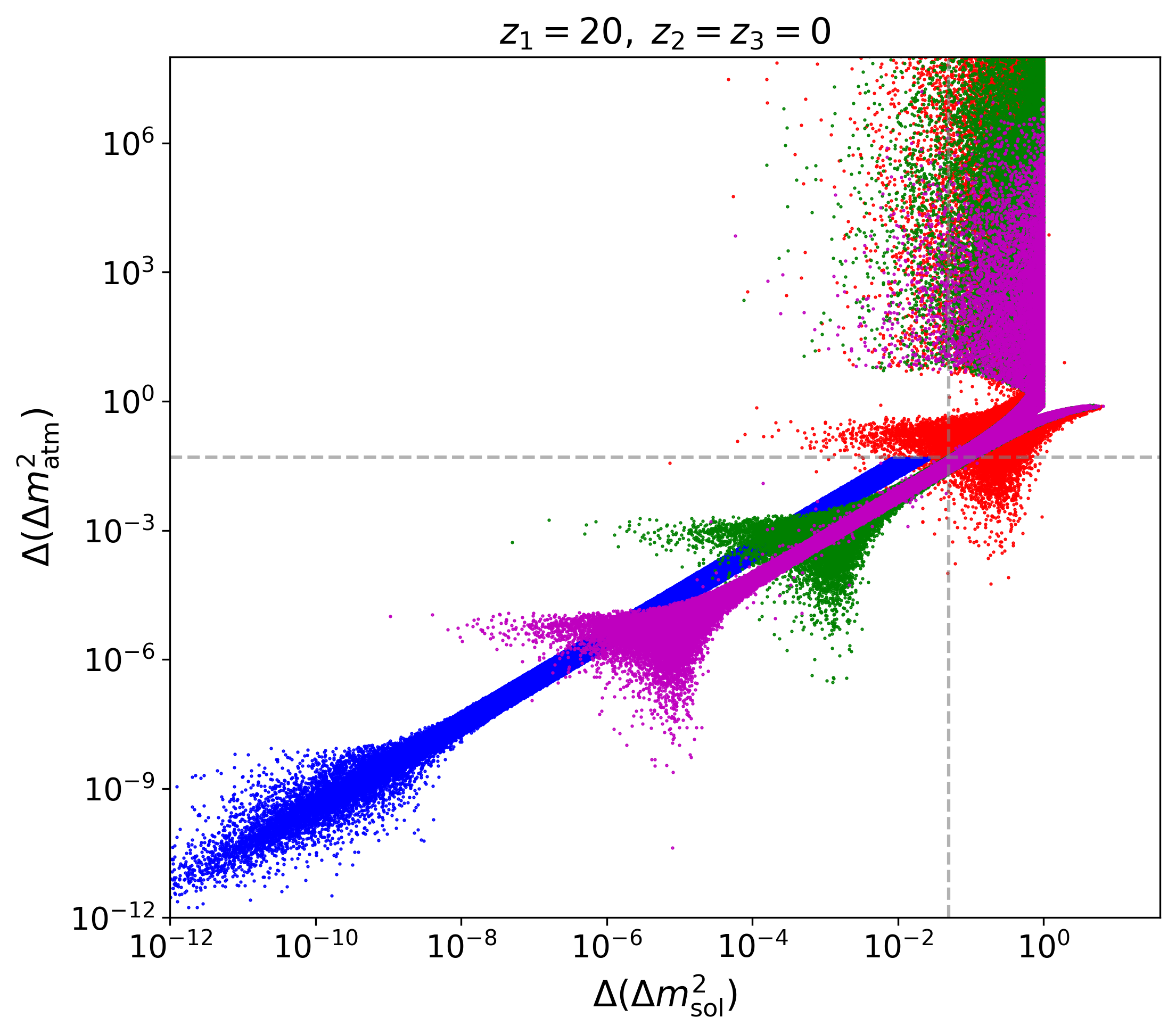}
\caption{Variation of the neutrino mass-squared differences obtained using 2- and 3-loop corrections, see text. }\label{fig111}
\end{figure}

In Figure~(\ref{fig111}), we show the deviation of the
reconstructed mass–squared differences with respect to the
experimental best–fit values when the two and three–loop
suppressed contributions are included. In particular, the red points correspond to the results obtained from the diagonalization of $\mathcal{M}^{\text{(1)}}$, the green points to $\mathcal{M}^{\text{(2)}}$ and the magenta points to $\mathcal{M}^{\text{(3)}}$. The results obtained from the diagonalization of the corresponding 
$\mathcal{M'}^{\text{(n)}}$ matrices are instead shown in blue.

As expected from the loop expansion, the higher–order terms
are strongly suppressed by powers of the loop factor
$(16\pi^2)^{-1}$ and therefore only marginally modify the reconstructed
neutrino mass spectrum. The numerical results
confirm that the perturbative expansion remains well under
control, and that the inclusion of higher–order corrections
does not significantly affect the phenomenological conclusions
of the present analysis.

\bibliography{bibliography}

\end{document}